\documentclass[aps,prl,twocolumn,showpacs,showkeys,superscriptaddress]{revtex4-1}

\usepackage{graphicx}

\usepackage{amsmath}

\begin{document}


\title{Strain-induced Defect Superstructure on the SrTiO$_3$(110) Surface}




\author{Zhiming Wang}

\affiliation{Beijing National Laboratory for Condensed Matter Physics \& Institute of Physics, Chinese Academy of Sciences, Beijing 100190, P. R. China.}
\affiliation{Institute of Applied Physics, Vienna University of Technology, Wiedner Hauptstrasse 8-10/134, A-1040 Vienna, Austria.}

\author{Fengmiao Li}

\affiliation{Beijing National Laboratory for Condensed Matter Physics \& Institute of Physics, Chinese Academy of Sciences, Beijing 100190, P. R. China.}

\author{Sheng Meng}

\affiliation{Beijing National Laboratory for Condensed Matter Physics \& Institute of Physics, Chinese Academy of Sciences, Beijing 100190, P. R. China.}

\author{Jiandi Zhang}

\author{E. W. Plummer}

\affiliation{Department of Physics and Astronomy, Louisiana State University, Baton Rouge, LA 70803, USA.}

\author{Ulrike Diebold}

\affiliation{Institute of Applied Physics, Vienna University of Technology, Wiedner Hauptstrasse 8-10/134, A-1040 Vienna, Austria.}

\author{Jiandong Guo}

\email{jdguo@iphy.ac.cn}

\affiliation{Beijing National Laboratory for Condensed Matter Physics \& Institute of Physics, Chinese Academy of Sciences, Beijing 100190, P. R. China.}



\begin{abstract}

We report on a combined scanning tunneling microscopy and density functional theory calculation study of the SrTiO$_3$(110)-(4~$\times$~1) surface. It is found that antiphase domains are formed along the [1$\bar{1}$0]-oriented stripes on the surface. The domain boundaries are decorated by defects pairs consisting of Ti$_2$O$_3$ vacancies and Sr adatoms, which relieve the residual stress. The formation energy of, and interactions between, vacancies result in a defect superstructure. It is suggested that the density and distributions of defects can be tuned by strain engineering, providing a flexible platform for the designed growth of complex oxide materials.

\end{abstract}

\maketitle

Controlling the structure is an effective way to tune the physical properties of materials. The structure of a bulk-synthesized material, however, is normally dictated by thermodynamics. On the surface, the structure of an epitaxial material can be adjusted by controlling both the growth dynamics and kinetics \cite{surfacegrow}. For example, surface strain controls the density of islands grown on conventional semiconductors, which is used for the production of quantum dot lasers \cite{dotreview,lasercorp}. Importantly, surface defects and their distribution strongly influence the growth and consequently the structure of the epitaxial material as well as its stability. A thorough knowledge of conventional semiconductor surface structures at the atomic scale has contributed greatly to the progress in electronic devices \cite{Kroemer01RMP}. In order to realize the promise of all-oxide electronics \cite{Mannhart10Science, Hwang10Science}, similar understanding and control of defects at the atomic scale must be achieved. This is challenging because, in addition to strain, oxides exhibit mixed valences of metal cations, surface polarity, nonstoichiometry, and complicated reconstructions \cite{Diebold03SSR, Noguera00JPCM, Castell08PRB, Moore07Sicence}.

Strontium titanate (SrTiO$_3$), a prototype perovskite oxide, has attracted extensive interest \cite{Ohtomo:2004uv, SantanderSyro:2010hf, Meevasana:2011ut, Ariando13NC, Herranz12SR}. The recently-resolved ($l$~$\times$~1) ($l$~= 3~-~6) series of reconstructions on the SrTiO$_3$(110) surface \cite{Enterkin:2010ie, Zhiming Wang PRL11} provides an opportunity to study complex structure of a polar surface at the atomic scale. The crystal can be considered as a stack of equidistant (SrTiO)$^{4+}$ and (O$_2$)$^{4-}$ planes along [110]. The bulk-truncated (110) surface is thus a polar surface with a structural instability \cite{Noguera00JPCM}. Furthermore, the ($l$~$\times$~1) series of reconstructions can be tuned by varying the surface stoichiometry \cite{Zhiming Wang PRB11, Zhiming Wang APL12}. The most commonly observed reconstruction has a (4~$\times$~1) symmetry [see Fig. 1(a)]. This structure consists of a layer of TiO$_4$ tetrahedra residing directly on the last (SrTiO)$^{4+}$ plane \cite{Enterkin:2010ie, Zhiming Wang PRL11}. The tetrahedra share oxygen corners, forming a network of six- and ten-membered rings. With a (Ti$_{6/4}$O$_{16/4}$) stoichiometry and formal charge of 2$^{-}$ per (1~$\times$~1) area, this added layer compensates the polarity of the (4~$\times$~1) surface \cite{Enterkin:2010ie}.

\begin {figure}[b]
 \includegraphics [width=3.4 in,clip] {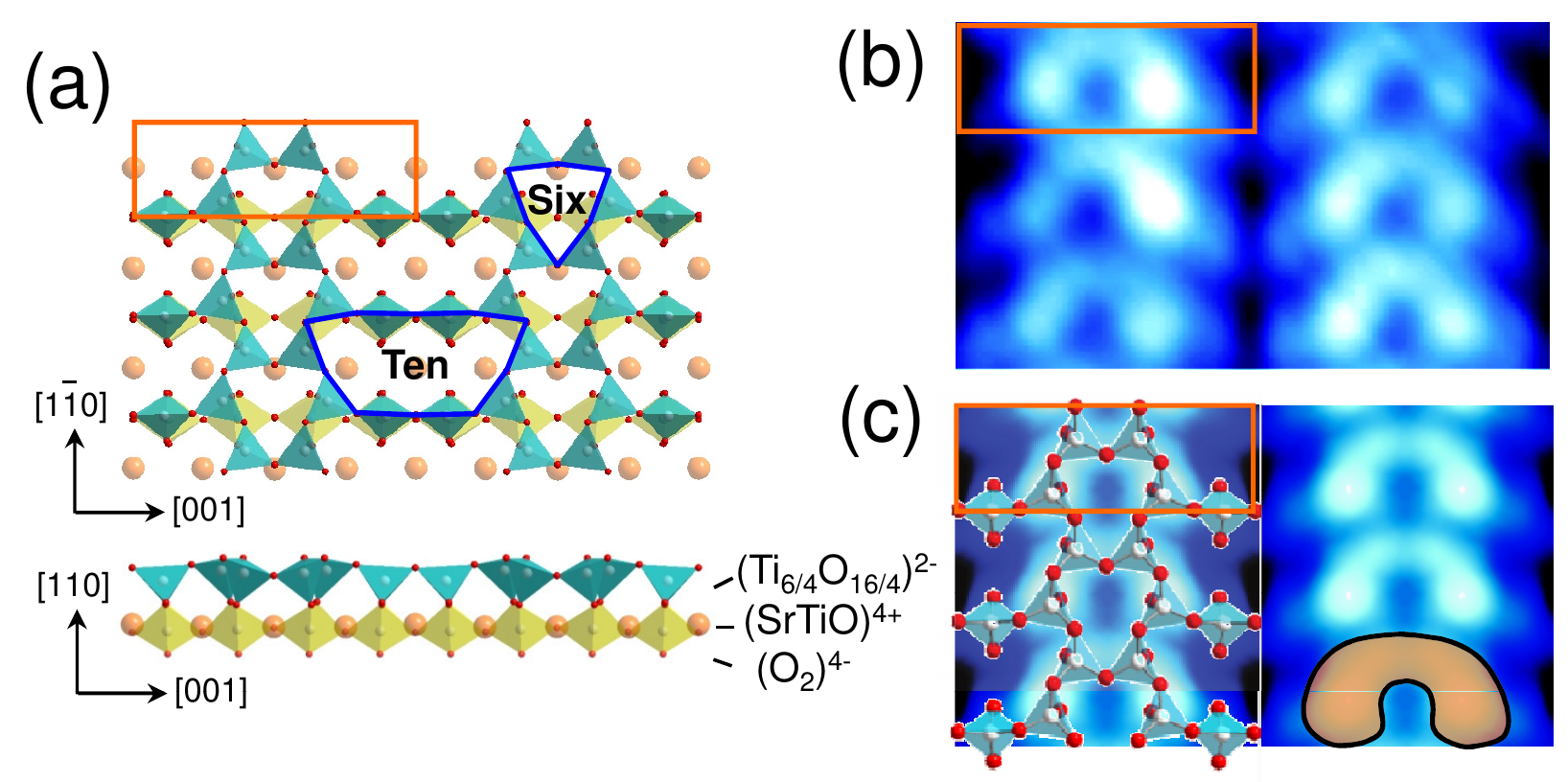}
 \caption{
(Color online) (a) Ball model of the SrTiO$_3$(110)-(4~$\times$~1) surface: (upper panel) top view and (lower panel) side view. The unit cell is marked by the rectangle. The structure consists of one layer of TiO$_4$ tetrahedra (green) on top of the SrTiO$_3$ lattice that contains TiO$_6$ octahedra (yellow). Both six- and ten-membered rings of corner-sharing tetrahedra are marked by blue polygons. (b) Experimental and (c) simulated STM topographic images of the (4$\times$1) surface. The structural model \cite{Zhiming Wang PRL11} is superimposed on (c). The higher tetrahedra in the six-membered rings give rise to a chain of bright, ``boomerang''-like features.
}
\end{figure}

\begin {figure*}[t]
 \includegraphics [width=5 in,clip] {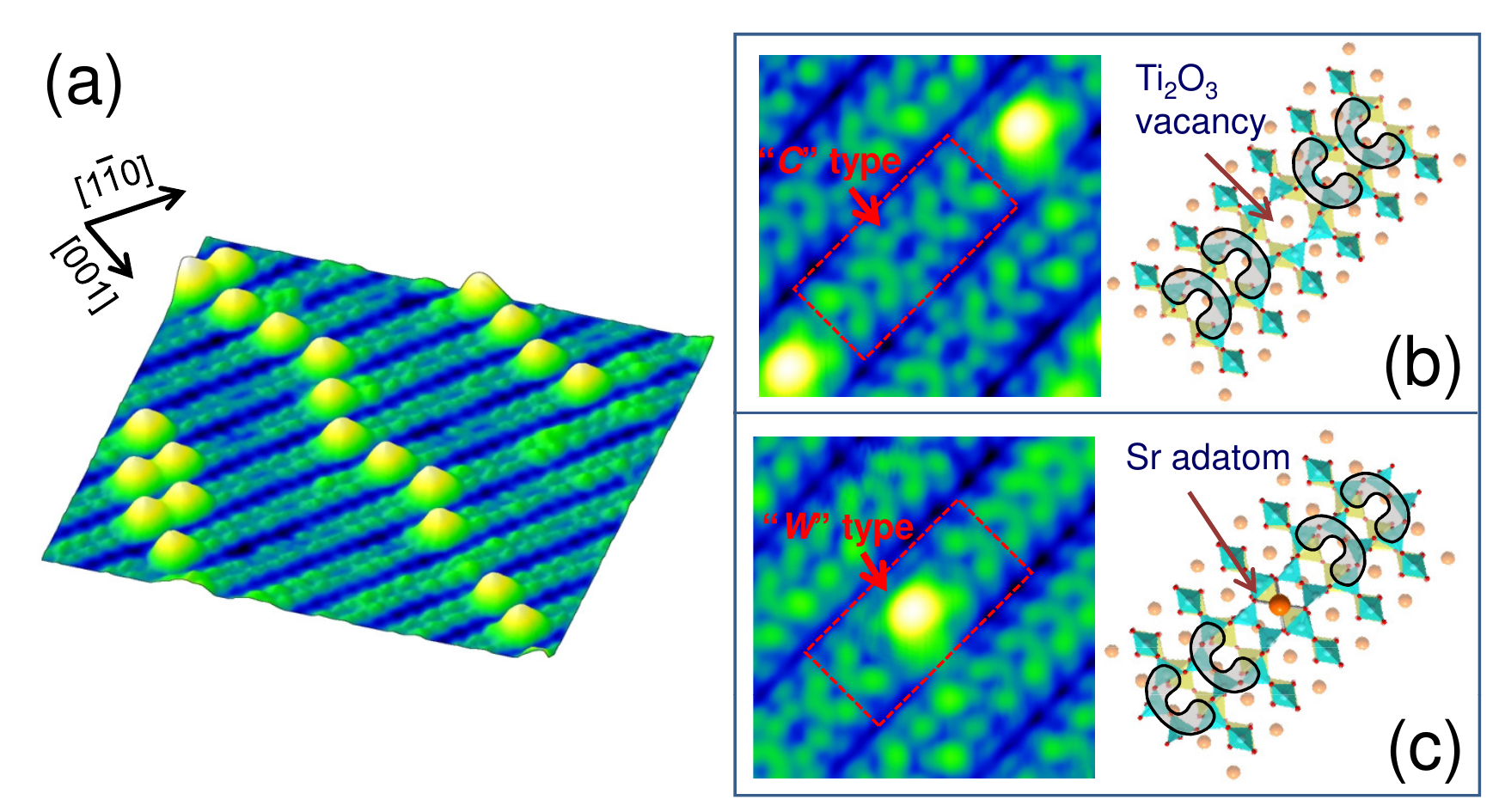}
 \caption{
(Color online) (a) STM image (16~$\times$~16 nm$^2$, +1.2 V/20 pA) of the SrTiO$_3$(110) surface. (b) and (c) STM images (the left panels, 5~$\times$~5 nm$^2$, +2 V/100 pA) and corresponding structural models of the ``C''-type and ``W''-type domain boundaries, with the wings of ``boomerangs'' pointing against and towards each other, respectively. A Ti$_2$O$_3$ cluster is removed from the C-type boundary while a Sr adatom is added to the W-type boundary.
}
\end{figure*}

In this Letter, we show that antiphase domains are formed on the SrTiO$_3$(110)-(4~$\times$~1) surface along the [1$\bar{1}$0]-oriented stripes. Sr adatoms and Ti$_2$O$_3$ vacancies appear as defect pairs decorating the domain boundaries. Those defect pairs relieve the residual stress on the surface and preserve the polarity compensation, while the interactions between the defects induce a quasi-ordered defect superstructure.

The scanning tunneling microscopy (STM) experiments were performed in the system with base pressure of 1~$\times$~10$^{-10}$ mbar at room temperature. The surface of Nb-doped (0.7 wt\%) STO(110) single crystal (12~$\times$~3~$\times$~0.5 mm$^3$) was cleaned by cycles of Ar$^+$ sputtering followed by annealing at 1000~$^{\circ}$C for 1~h \cite{Zhiming Wang APL09}. The annealing was carried out either in ultra-high vacuum (UHV) or in O$_2$ with the partial pressure up to 2$\times$10$^{-6}$~mbar without any difference observed on the surface. Density functional theory (DFT) calculations were carried out with the Vienna \textit{ab initio} Simulation Package (VASP) code \cite{VASP} with projector augmented-wave potentials \cite{PAW}. SrTiO$_3$(110)-(4~$\times$~$n$) ($n$~=~5~-~10, which refers to the periodicity of the quasi-ordered defect superstructure) were modeled with a supercell symmetrical along the [110] direction, consisting of a 9-layer slab separated by a vacuum layer of 12 \AA{} (for details, see Supplemental S3). Simulated STM images were obtained with Tersoff-Hamann approximation \cite{TH} by integrating the local density of empty states between Fermi level and 1.5~eV above the conduction band minimum. The (4~$\times$~$n$) surface stress and surface energy calculations were based on the stress theorem \cite{Nielsen83PRL,Kamisaka07SS} and the method described in literatures \cite{Noguera03PRB, Marks11PRL}, respectively.

Figure~1~(b) shows an unoccupied state STM topographic image of the SrTiO$_3$(110) surface, appearing as quasi-one-dimensional (1D) periodic chains along [1$\overline{1}$0] \cite{Zhiming Wang PRL11, Zhiming Wang PRB11}. Note that, while the bulk-terminated (110) surface has a mirror plane parallel to [001] (perpendicular to the rows/stripes), the six- and ten-membered rings of the added (4~$\times$~1) layer do not have this symmetry. This naturally leads to the formation of antiphase domain boundaries, as shown in Fig.~2. At one boundary [Fig.~2~(b)], the centers of the ``boomerangs'' point against each other (referred to as the C-type boundary), and a depression appears in the STM image. At the other boundary [Fig.~2~(c)], the wings of two neighboring boomerangs point against each other (referred to as the W-type boundary) where a bright dot is often found.

In order to match the two domains at a C-type domain boundary, one has to remove two tetrahedra, $\textit{i.e.}$, one Ti$_2$O$_3$ unit [see Fig.~2~(b) and Supplemental Fig.~S1]. The appearance of the resulting vacancy in simulated STM image based on the relaxed structure in DFT agrees well with the experimental one. Matching two domains at the W-type boundary results in a symmetric, 6-membered ring. This boundary is decorated by a single Sr adatom as verified by depositing additional Sr adatoms onto the surface (see Supplemental Fig.~S2), and by simulating STM image with DFT.

\begin {figure*}[t]
 \includegraphics [width=6.0in,clip] {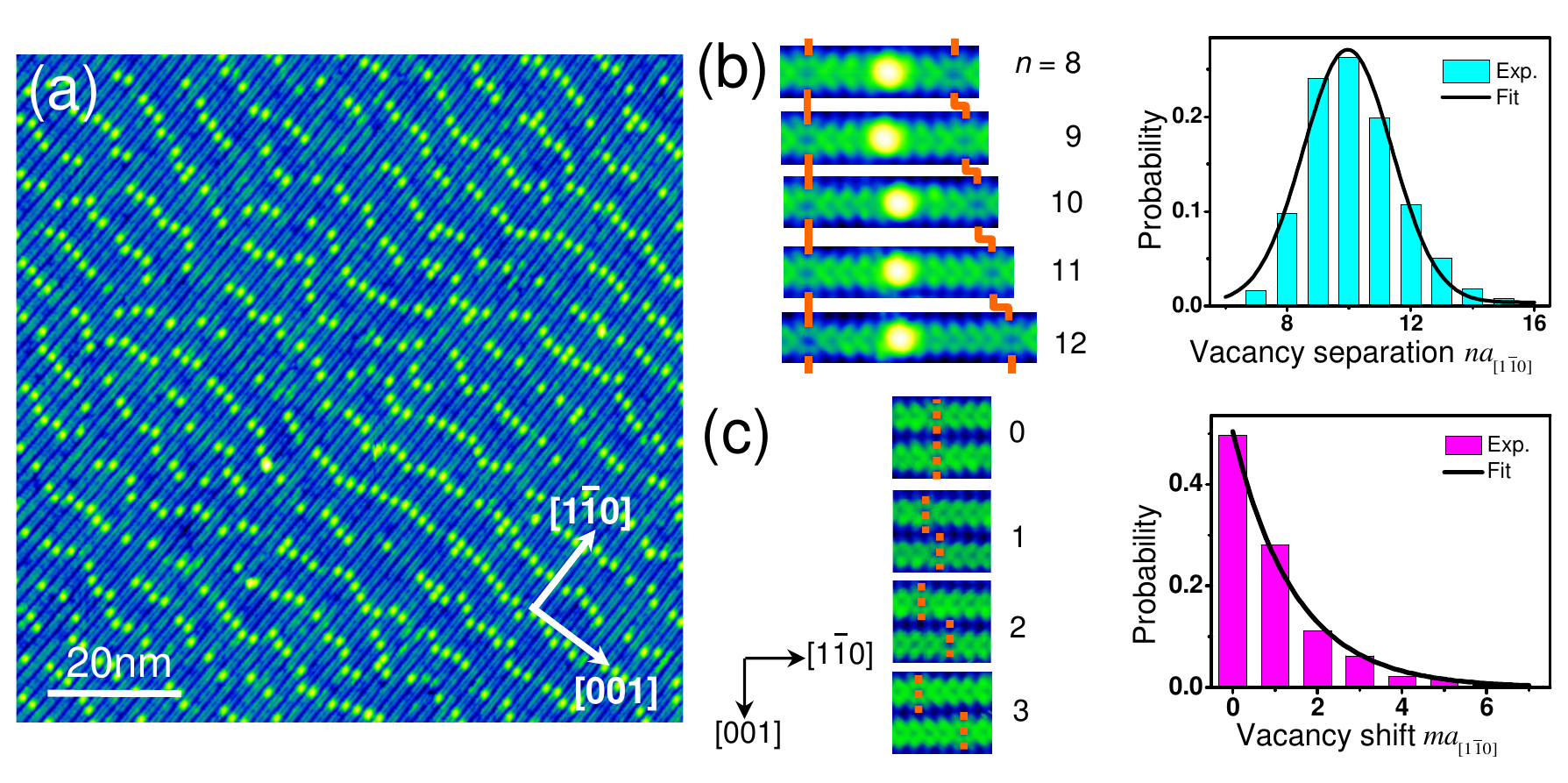}
 \caption{
(Color online) (a) Large scale STM image (100~$\times$~100 nm$^2$, 1.2 V/20 pA) of the SrTiO$_3$(110) surface.  (b) Separation ($n$) between two neighboring Ti$_2$O$_3$ vacancies at the C-type domain boundaries along [1$\overline{1}$0]. $n$ shows a Gaussian-type distribution. (c) Relative position of Ti$_2$O$_3$ vacancies in adjacent rows. The distribution of shift $m$ along the rows decreases exponentially.
}
\end{figure*}

The (4~$\times$~1) reconstruction compensates the surface polarity of SrTiO$_3$(110) \cite{Enterkin:2010ie}. A Ti$_2$O$_3$ vacancy at the type C boundary formally introduces two negative charges into the structure, thus disturbing the charge balance. This is compensated by a Sr adatom with two positive charges at the type W boundary. In other words, when the vacancies and adatoms appear as defect pairs, the overall polarity compensation on the (110) surface is maintained. This simple consideration is supported by the calculations of Bader charges \cite{Tang09JPCM} (Supplemental Table S1), which show that the nominal valences of Sr and Ti on the surface are the same as in the bulk, respectively.

Generally the formation of antiphase domain boundaries is accompanied by an energy cost. In equilibrium systems it is expected that both phases would condense into larger domains and the boundaries would tend to disappear. If the ``ideal'' (4~$\times$~1) phase (single domain without antiphase domain boundaries or defects) was the ground state of the SrTiO$_3$(110) surface, we should have been able to reduce the density of the domain boundaries (the defect pairs accordingly) by, $\textit{e.g.}$, prolonged annealing. However, we find domain boundaries with an unchanged density of $\sim$~0.09/nm$^2$ and a uniform distribution throughout the surface [Fig.~3~(a)] when we heat the sample to various temperatures up to 1200~$^{\circ}$C. On the other hand, depositing additional Sr adatoms on the surface (Supplemental Fig. S2) does not induce extra W-type domain boundaries. It is evident that the domain boundaries decorated by vacancies or adatoms are intrinsic to the (4~$\times$~1) surface, $\textit{i.e.}$, the ``ideal'' (4~$\times$~1) is not the ground state.

Interestingly, the distribution of defect pairs shows a stable, quasi-long-range ordering as shown in Fig.~3~(a). The separations between two neighboring vacancies within a [1$\overline{1}$0]-oriented stripe show a narrow distribution, with the favored value of $\sim$10 times of the lattice constant along [1$\overline{1}$0] ($a_{[1\overline{1}0]}$), as shown in Fig.~3~(b). Spatial correlations between vacancies and neighboring adatoms are also observed. High-resolution images in the left panel of Fig.~3~(b) show that the adatoms are located right at the center between two adjacent vacancies that are separated by $na_{[1\overline{1}0]}$ with $n$ = odd, and 1/2$a_{[1\overline{1}0]}$ away from the center for $n$ = even (see Supplemental S4).

The locations of vacancies in adjacent (4~$\times$~$n$) rows are also correlated. The statistics of their separation are evaluated in Fig.~3~(c); the probability of the relative shift along the stripe direction, $ma_{[1\overline{1}0]}$, shows an exponential decrease with $m$. Thus adjacent vacancies tend to align along [001]. Since Sr adatoms occupy the position at the center (or 1/2$a_{[1\overline{1}0]}$ away from the center) between two neighboring vacancies, they show a similar distribution along [001]. As a consequence, the vacancy-adatom pairs are assembled in meandering lines along [001], with roughly identical separations of 10$\times$$a_{[1\overline{1}0]}$. The ground state of the (4~$\times$~1) surface can be described as a superstructure of defect pairs with a (4~$\times$~10) periodicity.

In order to understand the energetics of the quasi-(4~$\times$~10) superstructure, DFT calculations were performed. We  first focus on the [1$\overline{1}$0] direction and find that an ``ideal'' (4~$\times$~1) surface is under stress. Within the top layer, the length of surface Ti-O bonds along [1$\overline{1}$0] ranges from $\sim$1.81~\AA{} to 1.85~\AA{}. In comparison, the calculated value for the SrTiO$_3$ bulk is 1.972~\AA{}, $\textit{i.e.}$, the bonds at the ``ideal'' (4~$\times$~1) surface are compressed by $\sim$~6\%. Quantitative surface stress calculations yield a residual compressive stress of -2.51~N/m along [1$\overline{1}$0]. We also evaluated the surface stress for various (4~$\times$~$n$) superstructures, with vacancy-vacancy separation of $na_{[1\overline{1}0]}$. As shown in Fig.~4~(a), the surface stress changes with $n$ and a switch from compressed to tensile at $n$~=~5~$\sim$~6 is estimated (for detail, see Supplemental S5). The surface stress relief is mainly achieved by the existence of vacancies at C-type boundaries (more than 90\% as shown in Supplemental Table S3), while the Sr adatoms at the W-type boundaries are responsible for the charge compensation. In the following analysis, we thus focus on the vacancies.

Considering surface stress only, the expected vacancy distribution does not agree with the experimental one. The experimental statistics peak at a vacancy-vacancy separation of $n$~=~10, as shown in Fig.~3~(b), while we estimate that the surface stress is completely relieved at a calculated value of $n$~\textless~6 [Fig.~4~(a)]. In addition to surface stress, however, other factors must be taken into account, such as the formation energies of defects and the interaction between them. Therefore we calculated the surface free energy of the (4~$\times$~$n$) superstructures. The results shown in Fig.~4~(b) indicate an energetically favorable defect-defect separation of $n$~=~9 or 10, in agreement with the experimental observations.

The DFT calculations use the (4~$\times$~$n$) superstructure, a 1D lattice with periodically-spaced vacancies. Thus the resulted surface energy includes the formation energy of each vacancies ($\Lambda$), as well as the interaction energy between vacancies ($E_{V-V}$). The energy difference between the (4~$\times$~$n$) superstructure and the ideal (4~$\times$~1) is given by:
\begin{equation}
\Delta E_S = 4na_{[001]}a_{[1\overline{1}0]}(\gamma_{4\times n}-\gamma_{4\times 1}) = \Lambda+E_{V-V}
\end{equation}
where a$_{[001]}$ is the lattice constant along [001], $\gamma_{4\times n}$ and $\gamma_{4\times 1}$ denote the surface energy (per unit cell) of (4~$\times$~$n$) and (4~$\times$~1), respectively. Since the vacancies are responsible for the surface stress relief, the interaction between them should be due to elastic forces. According to refs.~\cite{Lau77SS, Marchenko80JETP} $E_{V-V}$ depends on the spacing between adjacent vacancies ($na_{[1\overline{1}0]}$) as $E_{V-V}$=$G/n^3$ with $G$ a constant (independent on $n$) that characterizes the interaction strength. In the inset of Fig.~4~(b), $\Delta E_S$, the relative surface energy of (4~$\times$~$n$) is plotted as a function of $1/n^3$. It can be fitted to a straight line, consistent with the elastic force scenario. The positive slope reveals a repulsive interaction between vacancies \cite{Lau77SS, Marchenko80JETP}. The intercept of the line gives $\Lambda$~=~-0.426~eV; this is the energy gain per vacancy introduced at the C-type domain boundary.

\begin {figure}[t]
 \includegraphics [width=3.4in,clip] {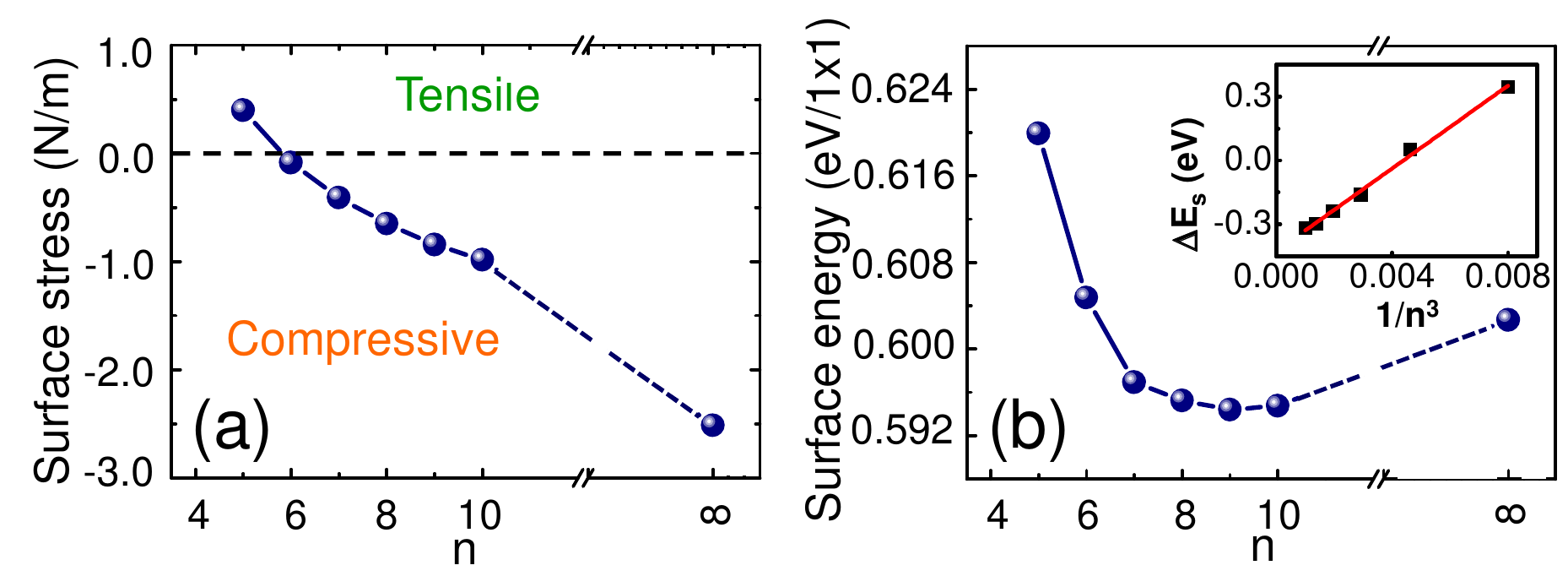}
 \caption{
(Color online) (a) Surface stress along the [1$\overline{1}$0] direction and (b) surface energy of (4~$\times~n$) surface. The inset indicates a linear relationship between the relative surface energy $\Delta E_S$ and 1/n$^3$. Note (4~$\times$~$\infty$) is equivalent to the ideal (4~$\times$~1) (with infinite defect separations).
}
\end{figure}

Equation 1 describes the competition between two mechanisms in the (4~$\times$~$n$) superstructure \textemdash ~introduction of (Ti$_2$O$_3$) vacancies lowers the surface energy by relieving the surface stress, while the repulsive interaction between them increases the surface energy. The former is proportional to $1/n$ (the vacancy density), while the later is proportional to $1/n^4$. Therefore, by minimizing $\gamma_{4\times n}$ as a function of $n$, we find that the most stable periodicity $n^{*} = (-4G/\Lambda)^{1/3} = 9.7$. Consistent with the experimental observations, such a value of $n^*$ corresponds to the separation of vacancies when the two competing mechanisms reach equilibrium.

It should be noted that the adatom-adatom interaction, $E_{A-A}$, and vacancy-adatom interaction, $E_{V-A}$, are neglected in the above analysis. Mediated by elastic force, $E_{A-A}$ is small since the Sr adatoms barely change the surface stress (see Supplemental Table~S3). Moreover, the electrostatic potential calculations with DFT show that the dipole interaction (including $E_{V-A}$) and coulomb repulsions are also small \cite{Dipole interactions}. Attributing the interactions between defects to mainly the vacancy-vacancy type provides a satisfactory approximation.

So far we have only considered 1D and repulsive interaction along [1$\overline{1}$0]. The formation of meandering lines [Fig. 3(a)] and the distribution of vacancy-vacancy shift in adjacent (4~$\times$~$n$) rows [Fig. 3(c)] suggest an attractive interaction along the perpendicular direction, [001]. The interaction energy can be determined quantitatively by either fitting the experimental statistics or calculating the total energies of appropriate modeled structures (Supplemental Fig.~S5). We estimate an attractive interaction energy of $\sim$0.22~eV for adjacent vacancies. The repulsive and attractive interactions along [1$\overline{1}$0] and [001], respectively, provide the essential driving forces for the formation of quasi-ordered (4~$\times$~10) defect superstructure.

The quasi-long-range ordered defects can serve as nucleation centers and guide the growth of an array of noble metal nanostructures with enhanced thermal stability \cite{Zhiqiang11JCP}. The regularly-distributed Sr adatoms may also act as effective dopants at the SrTiO$_3$-based heterointerfaces prepared by epitaxial growth on the anisotropic SrTiO$_3$(110) surface. More importantly, by applying a strain to the surface, $\textit{e.g.}$, by growing the SrTiO$_3$ template on an appropriate substrate, one can adjust the stress, thus changing the equilibrium between the vacancy-vacancy interaction and their formation energy. It is expected that a different superstructure would be obtained. Engineering the surface strain would allow for controlling the density/distribution of defects, providing the flexibility to tune the doping or to construct arrays of nanometric units for potential applications in oxide electronic devices.

In summary, we have investigated the formation mechanism of the (4~$\times$~10) defect superstructure on SrTiO$_3$(110) surface. The structural nonequivalence along [1$\overline{1}$0] of the (4~$\times$~1) surface leads to the intrinsic existence of antiphase domain boundaries. Defects (Ti$_2$O$_3$ vacancies and Sr adatoms) decorate the boundaries that are responsible for relieving the residual compressive stress due to lattice mismatch between the TiO$_4$ adlayer of the (4~$\times$~1) surface and the underlying bulk SrTiO$_3$. The repulsive elastic interaction between vacancies along [1$\overline{1}$0] that competes with their formation energy, as well as the attractive interaction of defects along [001], stabilizes the observed defect superstructure.

\begin{acknowledgments}

This work was supported by \textquotedblleft 973\textquotedblright~Program of China (2012CB921700) and NSFC Project 11225422. Z.W. and U.D. acknowledge partial support by the Austrian Science Fund (FWF) under Project No. F45, the ERC Advanced Research Grant 'OxideSurfaces', as well as helpful discussions with Michael Schmid and Cesare Franchini. J.Z. and E.P. are partially supported by US DOE DE-SC0002136.

\end{acknowledgments}


\begin{thebibliography}{}
\bibitem{Dipole interactions} The DFT calculations show that the modulation of the electrostatic potential around defects decays rapidly along [1$\overline{1}$0], suggesting that there are no significant dipole interactions.
\end{thebibliography}
\end{document}